\documentclass[11pt,twoside]{article}
\usepackage{amsmath}
\usepackage{amssymb}
\usepackage{amscd}
\usepackage{mathrsfs}
\usepackage{graphicx}
\usepackage{epstopdf}
\usepackage[usenames]{color}
\usepackage{wrapfig}
\usepackage{boxedminipage}
\newcommand{\cj}[1]{\overline{#1}}
\renewcommand{\.}{{\scriptstyle\boldsymbol{\dot{}}}}
\newcommand{\Lll}{{\scriptscriptstyle{\mathrm{L}}}}
\newcommand{\Rrr}{{\scriptscriptstyle{\mathrm{R}}}}
\newcommand{\into}{\hookrightarrow}
\newcommand{\onto}{\rightarrowtail}
\newcommand{\CC}{{\mathbb{C}}}
\newcommand{\LL}{{\mathbb{L}}}
\newcommand{\NN}{{\mathbb{N}}}
\newcommand{\RR}{{\mathbb{R}}}
\newcommand{\UU}{{\mathbb{U}}}
\newcommand{\id}{{1\!\!1}}
\newcommand{\we}{{\,\wedge\,}}
\newcommand{\weu}[1]{{\wedge^{\!#1}}}
\newcommand{\comp}{\mathbin{\raisebox{1pt}{$\scriptstyle\circ$}}}
\newcommand{\tn}{{\,{\otimes}\,}}
\newcommand{\ve}{{\,\vee\,}}
\newcommand{\Jcal}{{\mathcal{J}}}
\newcommand{\Lcal}{{\mathcal{L}}}
\newcommand{\Ucal}{{\mathcal{U}}}
\newcommand{\xx}{{\mathsf{x}}}
\newcommand{\zz}{{\mathsf{z}}}
\newcommand{\dx}{\dO\xx}
\newcommand{\bl}{{\bar\lambda}}
\newcommand{\bzz}{{\bar\zz}}
\newcommand{\be}{{\bar\varepsilon}}
\newcommand{\Bang}[1]{{\left\langle#1\right\rangle}}
\newcommand{\bang}[1]{{\langle#1\rangle}}
\newcommand{\de}{\partial}
\newcommand{\dde}[2]{\frac{\partial #1}{\partial #2}}
\newcommand{\na}{\nabla\!}
\newcommand{\nasl}{{\rlap{\raise1pt\hbox{\,/}}\nabla}}
\newcommand{\codiv}{\nabla\!{\cdot}}

\newcommand{\myitem}{\smallbreak\noindent$\bullet$}
\def\EndBox#1{\hskip0.1em\hfill\null\ \null\nobreak\hfill\kern3pt\hbox{$\scriptstyle #1$}\smallbreak}
\def\qed{\EndBox{\square}}
\renewcommand{\a}{\alpha}
\renewcommand{\b}{\beta}
\newcommand{\g}{\gamma}
\newcommand{\G}{\Gamma}
\renewcommand{\d}{\delta}
\newcommand{\e}{\varepsilon}
\renewcommand{\th}{\theta}
\renewcommand{\k}{\kappa}
\newcommand{\la}{\lambda}
\newcommand{\s}{\sigma}
\newcommand{\Ii}[2]{{}^{#1}_{\phantom{#1}\!#2}}
\newcommand{\iI}[2]{{}_{#1}^{\phantom{#1}\!#2}}
\newcommand{\iIi}[3]{{}_{#1\phantom{#2}\!\!#3}^{\phantom{#1}\!#2}}
\newcommand{\sA}{{\scriptscriptstyle A}}
\newcommand{\sB}{{\scriptscriptstyle B}}
\newcommand{\sC}{{\scriptscriptstyle C}}
\newcommand{\sD}{{\scriptscriptstyle D}}
\newcommand{\cA}{{\sA\.}}
\newcommand{\cB}{{\sB\.}}
\newcommand{\cC}{{\sC\.}}
\newcommand{\cD}{{\sD\.}}
\newcommand{\AAd}{{\sA\cA}}
\newcommand{\BBd}{{\sB\cB}}
\newcommand{\Ug}{\boldsymbol{\mathrm{U}}}
\newcommand{\SlG}{\boldsymbol{\mathrm{Sl}}}
\newcommand{\Lie}{\mathfrak{L}}
\newcommand{\E}{{\boldsymbol{E}}}
\newcommand{\F}{{\boldsymbol{F}}}
\newcommand{\Fa}{\cj{\F}{}^\lin}
\renewcommand{\H}{{\boldsymbol{H}}}
\newcommand{\M}{{\boldsymbol{M}}}
\newcommand{\U}{{\boldsymbol{U}}}
\newcommand{\Uc}{\cj{\U}}
\newcommand{\Ua}{\cj{\U}{}^\lin}
\newcommand{\Ul}{\U{}^\lin}
\newcommand{\V}{{\boldsymbol{V}}}
\newcommand{\Vc}{\cj{\V}}
\newcommand{\Va}{\cj{\V}{}^\lin}
\newcommand{\Vl}{\V{}^\lin}
\newcommand{\W}{{\boldsymbol{W}}}
\newcommand{\Wc}{\cj{\W}}
\newcommand{\Wa}{\cj{\W}{}^\lin}
\newcommand{\Wl}{\W{}^\lin}
\newcommand{\FL}{\F_{\!\!\Lll}} \newcommand{\FR}{\F_{\!\!\Rrr}}
\newcommand{\dO}{\mathrm{d}}
\newcommand{\HO}{\mathrm{H}}
\newcommand{\TO}{\mathrm{T}}
\newcommand{\TS}{\TO^{*}\!}
\newcommand{\eO}{\mathrm{e}}
\newcommand{\iO}{\mathrm{i}}
\newcommand{\Cs}{{\rlap{\lower3pt\hbox{\textnormal{\LARGE \char'040}}}{\Gamma}}{}}
\newcommand{\spec}[1]{{}_{\sst{\mathrm{#1}}}}
\newcommand{\sX}{{\scriptstyle X}}
\newcommand{\sY}{{\scriptstyle Y}}
\newcommand{\ssY}{{\scriptscriptstyle Y}}
\newcommand{\End}{\operatorname{End}}
\newcommand{\Tr}{\operatorname{Tr}}
\newcommand{\lin}{{\scriptscriptstyle\bigstar}}
\newcommand{\td}{\tilde}
\newcommand{\oh}{\tfrac{1}{2}}
\newcommand{\ih}{\tfrac{\iO}{2}}
\newcommand{\oq}{\tfrac{1}{4}}
\newcommand{\osq}{\tfrac{1}{\surd2}}
\newcommand{\mdots}{{\cdot}{\cdot}{\cdot}}
\newcommand{\sref}[1]{\S\ref{#1}}
\newcommand{\sst}{\scriptscriptstyle}

\def\EndBox#1{\hskip0.1em\hfill\null\ \null\nobreak\hfill\kern3pt\hbox{$\scriptstyle #1$}\smallbreak}
\def\qed{\EndBox{\square}}
\newtheorem{corollary}{Corollary}[section]

\newtheorem{proposition}{Proposition}[section]

\numberwithin{proposition}{section}

\numberwithin{corollary}{section}

\newcommand{\IP}{{\vphantom{I^I_I}}}
\newcommand{\sP}{{\scriptstyle P}}
\newcommand{\sV}{{\scriptstyle V}}
\newcommand{\chg}{\skew{4}\check\gamma}
\newcommand{\hag}{\skew{2}\hat\gamma}
\title{A first-order Lagrangian theory of fields with arbitrary spin}
\date{{\small January 4, 2018~(v2)} }
\author{{D.\ Canarutto} \\[6pt]
{\small\it Dipartimento di Matematica e Informatica ``U.~Dini'', }\\
{\small\it Via S. Marta 3, 50139 Firenze, Italia}\\
{\small email:~daniel.canarutto@unifi.it}\\
{\small http://www.dma.unifi.it/\char126 canarutto}}
\begin{document}

\maketitle

\begin{abstract}\noindent
The bundles suitable for a description of higher-spin fields
can be built in terms of a 2-spinor bundle as the basic `building block'.
This allows a clear, direct view of geometric constructions
aimed at a theory of such fields on a curved spacetime.
In particular, one recovers the Bargmann-Wigner equations
and the \hbox{$2(2j+1)$}-dimensional representation of the angular-momentum algebra
needed for the Joos-Weinberg equations.
Looking for a first-order Lagrangian field theory we argue,
through considerations related to the 2-spinor description of the Dirac map,
that the needed bundle must be a fibered direct sum
of a symmetric `main sector'|carrying an irreducible representation
of the angular-momentum algebra|and an induced sequence of `ghost sectors'.
Then one indeed gets a Lagrangian field theory that, at least formally,
can be expressed in a way similar to the Dirac theory.
In flat spacetime one gets plane-wave solutions that are characterised
by their values in the main sector.
Besides symmetric spinors, the above procedures can be adapted
to anti-symmetric spinors and to Hermitian spinors
(the latter describing integer-spin fields).
Through natural decompositions,
the case of a spin-2 field describing a possible deformation
of the spacetime metric can be treated in terms of the previous results.
\end{abstract}
\thispagestyle{empty}

\bigbreak\noindent

\bigbreak\noindent

\tableofcontents
\thispagestyle{empty}

\vfill\newpage

\subsubsection*{Introduction}

In comparison to the neatness of the Dirac theory of one-half-spin fields,
formulations of arbitrary-spin field theories suffer various complications~%
\cite{BargmannWigner48,Weinberg1964i,BarthChristensen83,Hurley1971,Hurley1972}.
Usual approaches proceed by considering fields
with many spinor and/or spacetime indices,
possibly constrained by symmetry conditions.
The ensuing angular-momentum representations
can turn out to be somewhat intricated.
Furthermore, matrix-based formalisms may tend to screen
the precise geometric role of the various involved objects by apparently putting
different operations on the same footing.

By contrast, we propose to examine the matter
in terms of direct, natural geometric constructions
performed by using the fundamental `building block' constituted by a
\emph{two-spinor space},
that is a 2-dimensional complex vector space endowed
with a certain algebraic structure.
This approach draws on a partly original treatment
of spinors and gauge field theories that has been explored
in previous papers~\cite{C98,C00b,C07,C14,C15b},
and is closely related|with differences we won't discuss in detail here|to
the Penrose-Reidler 2-spinor formalism~\cite{PR84,PR88,BarthChristensen83}.
One then avails of a direct description of higher-rank spinor spaces,
and redily grasps the working of the Dirac map
and of its extensions to such spaces.

In the above said context, we specially focus our attention on possible
first-order extensions of the Dirac theory.
We'll examine symmetric spinors
(\sref{ss:Symmetric spinors}-%
\ref{ss:Generalised algebraic Dirac equation}-%
\ref{ss:Generalised Dirac equation and plane waves}-%
\ref{ss:Lagrangian})
and other spinor types as well (\sref{ss:Further spinor field types}),
and comment about relations with some results found in the existing literature.

Formally, our treatment of spinors and spinor structures is
somewhat different from the prevailing approaches in the literature~%
\cite{BlaineLawsonMichelsohn89,DabrowskiPercacci86,FatibeneFrancaviglia98,Percacci86},
but we trust that the reader will promptly make the needed connections.

\subsection{Two-spinors and Dirac spinors}
\label{ss:Two-spinors and Dirac spinors}

In this and the next section we summarize our
approach to spinors and gauge field theories.

A finite-dimensional complex vector space $\V$ yields the associated
\emph{dual space} $\Vl$, \emph{anti-dual space} $\Va$
(that is the space of all anti-linear functions on $\V$)
and \emph{conjugate space} \hbox{$\Vc\equiv(\Va)^\lin$}.
There is a natural \emph{conjugation} anti-isomorphism
\hbox{$\Vl\to\Va:\la\mapsto\bar\la$} defined by
\hbox{$\bar\la(v)\equiv\overline{\la(v)}$}\,;
similarly, we have a conjugation anti-isomorphism \hbox{$\V\to\Vc$}.
A basis of $\V$ determines bases of the associated spaces.
Typically, `dotted indices' are used for the
components of elements in $\Vc$ and $\Va$.

The tensor product $\V{\otimes}\Vc$ has a natural \emph{real}
linear involution \hbox{$w\mapsto w^\dag$}
that is the composition of conjugation and tensor transposition.\footnote{
In terms of tensor components, the matrix $\smash{\bigl(w^\dag\bigr)}$ in any basis is
the complex-conjugated transposed matrix of $\smash{\bigl(w\bigr)}$,
or \hbox{$\smash{(w^\dag)^{\a\a\.}=\bar w^{\a\.\a}}$}
where `dotted indices'|as usual|denote components of conjugate spaces
and conjugation changes an index' type
(see also \S\ref{ss:Two-spinor soldering form and field theory}
for a brief remainder about that).}
Accordingly one gets the decomposition
$$\V{\otimes}\Vc=\HO(\V{\otimes}\Vc)\oplus\iO\HO(\V{\otimes}\Vc)$$
into the real eigenspaces,
respectively called \emph{Hermitian} and \emph{anti-Hermitian},
corresponding to involution eigenvalues $\pm1$\,.
This decomposition, applied to various cases, constitutes the main source for
the rich structure associated with spinor spaces.

The fundamental building block for our algebraic constructions
is a 2-dimensional complex vector space $\U$ such that $\weu2\U$
(not $\U$ itself) is equipped with a Hermitian structure;
in usual terms, the symmetry group of $\U$ is the complexified
special linear group $\SlG^c(2,\CC)$.
Thus we have a $\Ug(1)$-family of normalized complex symplectic forms
\hbox{$\e\in\weu2\Ul$},
but not one distinguished such object.
Up to a phase factor we then obtain isomorphisms
$$\e^\flat:\U\to\Ul:u\mapsto u^\flat\equiv\e(u,\_)~,\qquad
\e^\#:\Ul\to\U:\la\mapsto u^\#\equiv\e^\#(\la,\_)~,$$
where \hbox{$\e^\#\in\weu2\U$} is the inverse of $\e$\,.
We also obtain the conjugate isomorphisms \hbox{$\be^\flat:\Uc\to\Ua$}
and \hbox{$\be^\#:\Ua\to\Uc$}.

The 4-dimensional real vector space $$\H\equiv\HO(\U{\otimes}\,\Uc)$$
turns out to be naturally endowed with a Lorentzian metric $g$\,,
characterized by
$$g(u\tn\bar u\,,\,v\tn\bar v)=\e(u,v)\,\be(\bar u,\bar v)~.$$
The above operation
is actually independent of the phase factor affecting $\e$\,,
and the isotropic cone in $\H$ is constituted exactly by the elements
of the type $\pm u\tn\bar u$ with \hbox{$u\in\U$}
(thus one has a natural time orientation in $\H$:
future-pointing elements are characterized by the plus sign).
The dual space\footnote{
We indicate \emph{real duals} with an ordinary asterisk ($\square^*$),
and \emph{complex duals} with a star ($\square^\lin$).}
$\H^*$ can be naturally indentified with the
Hermitian subspace \hbox{$\HO(\Ul{\otimes}\,\Ua)$}.

Next we consider the 4-dimensional complex vector space
$$\W\equiv\U\oplus\Ua~,$$
which can be regarded as the space of \emph{Dirac spinors}.
Actually one gets a natural Clifford map
$$\g:\H\to\End\W~,$$
characterized by
$$\g[v\tn\bar v](u,\bar\la)=
\sqrt2\,\bigl(\bang{\bar\la,\bar v}\,v\,,\,\e(u,v)\,\bar v^\flat\bigr)~,\qquad
v\in\U\,,~
\bar v^\flat\equiv\be^\flat(\bar v,\_)\in\Ua~.$$
This operation is independent of phase factors in $\e$\,, too.

\begin{remark}
An element \hbox{$\sY\in\H\subset\U{\otimes}\,\Uc$} can be regarded as
a Hermitian scalar product on $\Ul$.
Similarly
$$\sY^\flat\equiv g^\flat(\sY)\in\H^*\cong\HO(\Ul{\otimes}\,\Ua)
\subset\Ul{\otimes}\,\Ua$$
can be regarded as a Hermitian scalar product on $\U$.
This may help to grasp the essential nature of $\g$\,.
Actually, observing that we have the restrictions\footnote{
Namely $\g[\sY]$ exchanges the (`chiral') subspaces \hbox{$\U,\Ua\subset\W$}.}
$$\smash{
\U~\overset{\g[{\sst Y}]}{{\longleftarrow}\!\!\!\!\!\!{\longrightarrow}}~\Ua}~,$$
we easily check that these
are exactly the linear maps \hbox{$\U\to\Ua$} and \hbox{$\Ua\to\U$}
determined by $\sqrt2\,\sY^\flat$ and $\sqrt2\,\sY$
regarded as Hermitian scalar products on $\U$ and $\Ul$, respectively.
Furthermore $\sY^\flat$ and $\sY$ are non-degenerate
iff \hbox{$g(\sY,\sY)\neq0$}\,,
and in that case $\sY^\flat/g(\sY,\sY)$
is the inverse metric of $\sY$.
If \hbox{$g(\sY,\sY)=1$} then $\sY^\flat$ and $\sY$
are inverse Hermitian metrics with signature $(+,+)$.
\end{remark}\medbreak

The space $\W$ is also naturally endowed with a Hermitian structure
with signature $(2,2)$ that is associated with the
anti-linear operation
$$\W\to\Wl:(u,\bar\la)\mapsto(\la,\bar u)~.$$
This is called the \emph{Dirac adjunction},
denoted by \hbox{$\psi\mapsto\bar\psi$} in the usual 4-spinor formalism.
The anti-linear mapping usually denoted as \hbox{$\psi\mapsto\psi^\dag$},
on the other hand,
is related to a \emph{positive} Hermitian structure which is
associated with the choice of an \emph{observer}|that is
a future-pointing time-like element in $\H$.

Besides $\U$ we'll assume a \emph{unit space} $\LL$\,,
that is a real 1-dimensional semi-vector space,
regarded as the space \emph{of length units}.\footnote{
A convenient, general setting for the treatment of physical units
(see~\cite{JMV10,C12a} for details)
was introduced after an idea by M.~Modugno and then adopted
by several authors.
The consistent use of that approach is indeed
a source of mathematical clarity, and as such it is used here
(it is \emph{not} specially needed for higher-spin fields).
A \emph{unit space} is defined to be a 1-dimensional real semi-space,
namely a positive semi-field associated with the semi-ring $\smash{\RR^+}$.
The \emph{square root} $\smash{\UU^{1/2}}$ of a unit space $\UU$
is defined by the condition that $\smash{\UU^{1/2}\tn\UU^{1/2}}$ be isomorphic to $\UU$.
More generally, any \emph{rational power} of a unit space
is defined up to isomorphism,
with negative powers corresponding to dual spaces.
Essentially,
in this paper we only use the unit space $\LL$ of lengths and its powers;
this amounts to making the common choice \hbox{$\hbar=c=1$}\,.}
We remark that rational powers of unit spaces are naturally defined;
integer powers, in particular, are tensor powers,
and negative powers denote dual spaces.
Moreover we stress that both $\U$ and $\LL$ can be derived from the unique
`algebraic datum' constituted by a 2-dimensional complex vector space
with no added assumptions|that derivation however needs some
extra constructions that are not actually used here.

\subsection{Two-spinor soldering form (tetrad) and field theory}
\label{ss:Two-spinor soldering form and field theory}

Next we consider a 4-dimensional real manifold $\M$
and a complex vector bundle \hbox{$\U\onto\M$} whose fibers are endowed
with the structure described in~\sref{ss:Two-spinors and Dirac spinors},
as well as the induced bundles \hbox{$\H\onto\M$} and \hbox{$\W\onto\M$}.
In this setting we consider the following fields.

\myitem~The Dirac field is a section \hbox{$\psi:\M\to\LL^{-3/2}\tn\W$};
the adjoint Dirac field is a section \hbox{$\bar\psi:\M\to\LL^{-3/2}\tn\Wl$}
that in general can be regarded as independent from $\psi$\,,
though eventually the field equations are mutually Dirac-adjoint.

\myitem~The tetrad field
is described as a section \hbox{$\th:\M\to\LL\tn\TS\M\tn\H$},
and can be viewed as a linear morphism \hbox{$\TO\M\to\LL\tn\H$}.
A non-degenerate tetrad can be regarded as a \emph{soldering form},
bringing the fiber structure of $\H$ to $\TO\M$.
More precisely the Lorentz metric in the fibers of $\H$ determines,
through $\th$\,, an $\LL^2$-\emph{scaled}\footnote{
Scalar products have the physical dimension of a square length.}
Lorentz metric of $\M$,
denoted for simplicity still as $g$ and given by
$$g(\sX,\sY)\equiv g(\th(\sX),\th(\sX'))~,\qquad \sX,\sX'\in\TO\M~.$$
Furthermore a soldering form determines a \emph{scaled Dirac map}
$$\g:\TO\M\to\LL\tn\End\W:\sX\mapsto\g[\th(\sX)]~,\qquad \sX\in\TO\M~.$$

\myitem~Finally we consider a linear connection $\Cs$ of \hbox{$\U\onto\M$},
preserving the algebraic fiber structure of $\weu2\U$\,.
This \emph{2-spinor connection} naturally determines linear connections
of the related bundles $\Ul$, $\Ua$, $\Uc$, $\H$ and $\W$.
A couple $(\th,\Cs)$ determines a spacetime connection $\G$,
characterized by the condition that $\th$ itself be covariantly constant;
note that this $\G$,
which turns out to be metric but not necessarily torsion-free,
is to be regarded not as a fundamental field but rather as a byproduct.
The 2-spinor connection can be decomposed into a \emph{purely gravitational part}
and a \emph{gauge field}.
These points can be conveniently expressed in terms of components,
as we are going to do after introducing some further notational details.
\smallbreak

\begin{remark}
The notion of soldering form considered here is essentially equivalent
to the notion variously referred to in the literature~%
\cite{HehlHeydeKerlickNester76,Henneaux1978,Trautman06,Canarutto2018a}
as a `frame field', or a `tetrad' or a `vierbein'
(`vielbein' for arbitrary dimension).
A global soldering form, with the target bundle $\H$
constructed from the spinor bundle itself,
yields a `spin structure' on $\M$
(in this paper we are not concerned with the possible topological obstructions
to the existence of such structures).
We also note that the fact that $\H$
is explicitely constructed from the spinor bundle is important
in the study of gauge field theories coupled with gravity~%
\cite{C98,C00b,C07,C10a,C15b}.
\end{remark}\bigbreak

A local frame $\bigl(\zz_\sA\bigr)$ of $\U$, \hbox{$A=1,2$},
determines the dual frame $\bigl(\zz^\sA\bigr)$ of $\Ul$,
the conjugate frame $\bigl(\bzz_\cA\bigr)$ of $\Uc$
and the anti-dual frame $\bigl(\bzz^\cA\,\bigr)$ of $\Ua$.
We will only consider special such frames,
that are characterized by the condition that
\hbox{$\e=\e_{\sA\sB}\,\zz^\sA\we\zz^\sB$}
is a normalized section \hbox{$\M\to\weu2\Ul$}
(where \hbox{$\e_{\sA\sB}\equiv\d^1_\sA\d^2_\sB-\d^2_\sA\d^1_\sB$}).
We'll use shorthands
$$u_\sA\equiv(u^\flat)_\sA=\e_{\sB\sA}\,u^\sB~,\qquad
\la^\sA\equiv(\la^\#)^\sA=\e^{\sB\sA}\,\la\sB~,\qquad
u\in\U\,,~\la\in\Ul\,,$$
as well as analogous conjugate shorthands.
The isomorphism \hbox{$\H\to\H^*$} associated with the Lorentz metric
can now be similarly expressed as
$$\sY\!\!_\AAd\equiv(g^\flat(\sY))_\AAd=\e_{\sB\sA}\,\be_{\cB\cA}\,\sY^\BBd~,\qquad
\sY\in\H~.$$

We also consider the induced \emph{Pauli frame}
$\bigl(\tau_\la\bigr)$ of $\H$, \hbox{$\la=0,1,2,3$}, where
$$\tau_\la\equiv\osq\,\sigma\iI\la\AAd\,\zz_\sA\tn\bzz_\cA\in\H\subset\U{\otimes}\,\Uc$$
is written in terms of the Pauli matrices $\sigma_\la$\,.
This frame turns out to be orthonormal.
Let moreover $\bigl(\xx^a\bigr)$ be a local coordinate chart of $\M$.
We obtain the coordinate expressions
\begin{align*}
&\th=\th_a^\la\,\dx^a\tn\tau_\la=\th_a^\AAd\,\dx^a\tn\zz_\sA\tn\bzz_\cA~,\qquad
\th_a^\la\,,\th_a^\AAd:\M\to\RR\tn\LL~,
\\[6pt]
&\g[\sX](u,\bar\la)\equiv\sqrt2\,\sX^a\,\th_a^\AAd\,\bigl(
\bar\la_\cA\,\zz_\sA~,~\e_{\sB\sA}\,\be_{\cB\cA}\,u^\sB\,\bzz^\cB\,\bigr)~,
\\[6pt]
&g_{ab}=g_{\la\mu}\,\th_a^\la\,\th_b^\mu~.
\end{align*}

Let now $\Cs\!\iIi a\sA\sB$ be the components of a 2-spinor connection.
The induced connection of $\weu2\U$ has the components
\hbox{$\hat\Cs\!_a=\Cs\!\iIi a\sA\sA$}\,,
hence the condition that $\Cs$ preserves the Hermitian structure of $\weu2\U$
can be expressed as the requirement that these components are imaginary, namely
$$\hat\Cs\!_a=2\,\iO\,{\mathrm{B}}_a~,\qquad {\mathrm{B}}_a:\M\to\RR~.$$
The components of the induced connection $\td\G$ of \hbox{$\H\subset\U{\otimes}\,\Uc$}
can be expressed as
$$\td\G\!\iIi a\AAd\BBd=
\Cs\!\iIi a\sA\sB\,\d\Ii\cA\cB+\d\Ii\sA\sB\,\bar\Cs\!\iIi a\cA\cB~.$$
These turn out to be traceless,
and can be seen as characterizing the `gravitational part' $\td\Cs$ of $\Cs$.
Actually by straightforward computations one gets
$$\Cs\!\iIi a\sA\sB=\iO\,{\mathrm{B}}_a\,\d\Ii\sA\sB+\td\Cs\!\iIi a\sA\sB\equiv
\oh\,\Cs\!\iIi a\sC\sC\,\d\Ii\sA\sB+\oh\,\td\G\!\iIi a{\sA\cA}{\sB\cA}~.$$

We can also write the components of the induced
\emph{4-spinor connection} of $\W$ as
$$\Cs\!\iIi a\a\b=
\iO\,{\mathrm{B}}_a\,\d\Ii\a\b
+\tfrac{1}{8}\,\td\G\!\iIi a\la\mu\,(\g_\la\,\g^\mu-\g^\mu\,\g_\la)\Ii\a\b~,\qquad
\g_\la\equiv\g[\tau_\la]~,$$
where we used the components of $\td\G$
in the Pauli frame $\bigl(\tau_\la\bigr)$ associated with $\bigl(\zz_\sA\bigr)$\,.

A Lagrangian theory of the fields $(\psi,\bar\psi,\th,\Cs)$
can be formulated by writing down a straightforward translation
of usual Lagrangian densities in terms of the above described
formalism~\cite{C98,C00b}.
One gets essentially the standard field equations.
In particular one gets the Dirac equations
$$\begin{cases}
-\iO\,\nasl\bar\psi-m\bar\psi-\ih\bar\psi\,\g[\breve T]=0~,
\\[6pt] \phantom{-}
\iO\,\nasl\psi-m\psi-\ih\g[\breve T]\psi=0~,
\end{cases}$$
where \hbox{$\nasl\psi\equiv g^{ab}\,\g_a\na_b\psi$}\,,
\hbox{$\nasl\bar\psi\equiv g^{ab}\,\na_b\bar\psi\comp\g_a$}\,,
and \hbox{$\breve T=\breve T_a\,\dx^a\equiv T\Ii{b}{ab}\,\dx^a$}
is the 1-form naturally associated with the torsion.

A non-Abelian version of the above sketched theory can be obtained by replacing
the bundle $\U$ with \hbox{$\U'\equiv\U\tn\F$},
where the bundle \hbox{$\F\onto\M$} is endowed with a Hermitian fiber structure.
With no loss of generality, the gauge part
of the considered connection $\Cs'$ of $\U'$
can be completely attributed to a linear Hermitian connection $\k$ of $\F$,
leaving the gravitational contribution associated with $\U$ only.
Even more generally one can assume different `right' and `left'
Hermitian bundles $\FR$ and $\FL$\,,
and define the Fermion bundle to be~\cite{C10a,C15b}
$$\W'\equiv(\FR\tn\U)\oplus(\FL\tn\Ua)~.$$

\subsection{Higher-spin extensions of the Dirac map}
\label{ss:Higher-spin extensions of the Dirac map}

In the literature, the notion of a field of arbitrary
integer or half-integer spin is usually introduced by adding
spacetime indices and/or spinor indices to the field's components.
All possibilities can be recovered by viewing the field under consideration
as a section of some sector of the tensor algebra bundle $\U^{\otimes}$
generated by the two-spinor bundle $\U$ and its associated bundles
$\Ul$, $\Uc$ and $\Ua$;
in particular, spacetime indices are related to the Hermitian subspaces
\hbox{$\H\subset\U{\otimes}\,\Uc$} and \hbox{$\H^*\subset\Ul{\otimes}\,\Ua$}.
Using this description, one readily determines the natural operations
that are relevant to our purpose,
without being involved with intricacies of matrix group representations.

The general idea is that a field of spin $j$ is described as a section
of some sector of tensor rank \hbox{$r\equiv2j\in\NN$}\,.
For each \hbox{$\sY\in\H$}, the Dirac map determines mutually transpose
linear morphisms
$$\U~\overset{\textstyle\g[\sY]}{{\longleftarrow}\!\!{\longrightarrow}}~\Ua~,
\qquad
\Ul~\overset{\textstyle\g[\sY]}{{\longleftarrow}\!{\longrightarrow}}~\Uc~,$$
that are denoted for simplicity by the same symbol.
We want to extend this action to an action on $\U^{\otimes}$.
For any sector of tensor rank $r$ we naturally obtain $r$ different extensions:
$$\g_{\sst(n)}\equiv\underbrace{\id\tn\mdots\tn\id}_{n-1~\text{factors}}
\tn\g[\sY]\tn\underbrace{\id\tn\mdots\tn\id}_{r-n~\text{factors}}~,\qquad
1\leq n\leq r~.$$
Thus we can introduce maps \hbox{$\H\to\End(\U^{\otimes})$}.
It should be noted that these, though natural, need not be Clifford maps.

One important point to take into account
is whether a certain chosen extension of $\g$
is valued into the endomorphisms of the sector under consideration.
This affects the type of field equation that can be introduced.
When $\g$ yields sector endomorphisms, we may consider a field equation
analogous to the Dirac equation, with a mass term.
Otherwise we must deal with a massless field.
The basic example of the latter is, obviously, the neutrino field,
which can be described as a section \hbox{$\M\to\Ua$}.

The same type of obstruction to massive fields obeying
an extension of the Dirac equation applies to fields with ``spacetime'' indices,
that is indices pertaining to \hbox{$\H\subset\U{\otimes}\,\Uc$}
or \hbox{$\H^*\subset\Ul{\otimes}\,\Ua$}
(see \sref{ss:Further spinor field types}).
Actually neither $\g_{\sst(1)}$ nor $\g_{\sst(2)}$
are valued into the endomorphisms of $\U{\otimes}\,\Uc$.

By contrast, the tensor algebra $\W^{\otimes}$
generated by \hbox{$\W\cong\Wa$} and \hbox{$\Wl\cong\Wc$}
is preserved under the action of any $\g_{\sst(n)}$\,,
as well as any sectors of its obtained by algebraic operations such as
symmetrization, anti-symmetrization and rank restriction.
One gets extensions \hbox{$\nasl\!_{\sst(n)}\equiv\g_{\sst(n)}^{\,a}\na_a$}
of the Dirac operator, acting on sections
of such \emph{chirally symmetric} sectors.
For a given chirally symmetric sector of tensor rank $r$
one can write down $r$ first-order equations
$$\pm\iO\nasl\!_{\sst(n)}\!\Psi=m\Psi~,\qquad 1\leq n\leq r~,$$
that are essentially the \emph{Bargmann-Wigner equations}.
In general there is no distinguished way to select one equation,
and the set of $r$ equations corresponds to no obvious, natural Lagrangian formulation~%
\cite{BargmannWigner48,Dvoeglazov2003,Kaparulin_etal2013,HuangRuanWuZheng2002,
Jeffery78,LorenteRodriguez1983}.

Alternatively one may consider a single equation of order $r$\,, namely
$$\pm\iO^r\nasl\!_{\sst(1)}\dots\nasl\!_{\sst(r)}\!\Psi=m^r\Psi~.$$
This is essentially the \emph{Joos-Weinberg equation}~%
\cite{Weinberg1964i,Weinberg1964ii,Weinberg1969,Joos1962,
Shay1968,DelgadoAcosta_etal2015}.

Our approach to fields of arbitrary spin will be somewhat different.
We claim that, starting from an arbitrary sector \hbox{$\E\subset\U^{\otimes}$},
we can obtain a theory with a first-degree field equation
by considering a suitable extension \hbox{$\skew4\tilde\E\supset\E$}
and a sufficient number of independent auxiliary `ghost' fields.
The essential requirement is that this extension be closed
under the extended action of $\g$\,.

A possible general form of the above said first-degree field theory
can be sketched as follows.
Let the extended bundle be of the type
\hbox{$\skew4\tilde\E\equiv\E\oplus\E'\oplus\E''\oplus\mdots$}\,,
and let \hbox{$\g',\g'',\dots$} be chosen among the $\g_{\sst(n)}$
in such a way that for any \hbox{$\sY\in\H$} one gets a sequence
$$\E\stackrel{\g'[\ssY]}{\longrightarrow}
\E'\stackrel{\g''[\ssY]}{\longrightarrow}\E''
\longrightarrow\mdots\longrightarrow\E~.$$
Then we have a morphism \hbox{$\skew3\td\g:\H\to\End\skew4\tilde\E$}
(not a Clifford map in general),
and that yields a `Dirac' operator
\hbox{$\smash{\tilde{\nasl}}\equiv\skew3\td\g^a\na_a$}
acting on sections \hbox{$\Psi:\M\to\skew4\tilde\E$}.
Accordingly we can write a `Dirac' equation
\hbox{$\iO\smash{\tilde{\nasl}}\Psi=m\Psi$}.
In the special case of flat spacetime this equation admits plane wave
solutions which are actually determined by their restrictions \hbox{$\M\to\E$}.

We'll call $\E$ the \emph{main sector} and $\E',\E''\dots$
the \emph{ghost sectors}.

Furthermore by considering an independent dual field
\hbox{$\bar\Psi:\M\to\skew4\tilde\E{}^*$}
we can write down the theory's Lagrangian
in a form similar to the usual Dirac Lagrangian
(e.g.\ see~\sref{ss:Lagrangian}).

\subsection{Symmetric spinors}
\label{ss:Symmetric spinors}

Symmetric spinors have a special status in the literature
about higher-spin fields.
Actually we will see that
the extensions of the Dirac map and of the Dirac operator
introduced in~\sref{ss:Higher-spin extensions of the Dirac map}
work most naturally in the symmetric and anti-symmetric cases.

Furthermore, symmetric spinors are special with regard to representations
of the angular-momentum Lie algebra $\Lie$,
usually treated as the matrix Lie algebra ${\boldsymbol{\mathfrak{su}}}(2)$\,.
$\Lie$ can be realized as the Lie subalgebra of $(\End\U,[,])$
consituted of all traceless endomorphisms that are anti-Hermitian
with respect to some Hermitian metric $h$ of $\U$.
Namely, the choice of $h$ determines a representation
\hbox{$\rho\equiv-\iO J:\Lie\into\End\U$},
where the Hermitian-valued map $J$ is simply called \emph{angular-momentum}.
On turn, this determines the representation
\hbox{$(\rho,-\bar\rho^*):\Lie\to\End\W$},
and the representations
$${\otimes}^r\rho\equiv\rho\tn\id\tn\mdots\tn\id
~+\dots+~\id\tn\mdots\tn\id\tn\rho~:~\Lie\to\End({\otimes}^r\U)~,\qquad
r\in\NN~.$$
By symmetric restrictions one also gets the representations
$$\rho^{\sst(r)}:\Lie\to\End({\vee}^r\U)~,$$
where $\vee$ denotes the symmetrized tensor product.

\begin{proposition}
Let $h$ be any Hermitian metric of $\U$ and \hbox{$\rho:\Lie\into\End\U$}
the representation induced by it.
Then $\rho^{\sst(r)}$ is an irreducible $r\,{+}\,1$-dimensional representation
for any \hbox{$r\in\NN$}\,.
\end{proposition}{\sc track of proof}.
An $h$-orthonormal basis $\bigl(\zz_\sA\bigr)$ of $\U$ also yields the basis
$$\bigl(\rho_i\bigr)\equiv\bigl(-\iO J\!_i\bigr)\equiv
\bigl(-\ih\,\s\iIi i\sA\sB\,\zz_\sA\tn\zz^\sB\bigr)\subset\Lie~,\qquad
i=1,2,3\,,$$
where $\s_i$ is the $i$-th Pauli matrix.
One readily checks that $\bigl(\rho_i\bigr)$
is an orthonormal basis with respect to the Euclidean metric
\hbox{$(A,B)\mapsto-2\Tr(A\comp B)$} of \hbox{$\Lie\subset\End\U$}.
A straightforward calculation then shows that the symmetric-valued
\hbox{$(J^{\sst(r)})^2\equiv
J^{\sst(r)}_{\!1}\!{\scriptstyle\circ}\,J^{\sst(r)}_{\!1}
+J^{\sst(r)}_{\!2}\!{\scriptstyle\circ}\,J^{\sst(r)}_{\!2}
+J^{\sst(r)}_{\!3}\!{\scriptstyle\circ}\,J^{\sst(r)}_{\!3}$}
is proportional to the identity, namely we obtain
$$(J^{\sst(r)})^2\phi=\bigl[\bigl(\tfrac r2\bigr)^2+\tfrac r2\bigr]\,\phi~,
\qquad \phi\in{\vee}^r\U~.$$
Since \hbox{$\dim{\vee}^r\U=r\,{+}\,1$}\,,
from basic results in representation theory one then finds that
\hbox{$\rho^{\sst(r)}$} is an irreducible
representation with ``total angular momentum quantum number''
\hbox{$j=r/2$}\,.\qed

The discussion in~\sref{ss:Higher-spin extensions of the Dirac map}
shows that the sector \hbox{${\vee}^r\U\equiv{\vee}^{2j}\U$} alone
is not suitable for a first-order field theory,
since it is not closed for any natural extension of the Dirac algebra.
Thus we are led to regard this sector as the `main sector' of such a theory,
associated with ghost sectors in such a way that
together they constitute a chirally symmetric
(\sref{ss:Higher-spin extensions of the Dirac map})
sub-bundle of the tensor algebra ${\otimes}^r\W$.
A natural decomposition of ${\vee}^r\W$ then comes to mind.
We first introduce the convenient shorthand
$$\U^{(h,k)}\equiv{\vee}^{h}\U\tn{\vee}^{k}\Ua~,\qquad
0\leq h,k\leq r~.$$
\begin{proposition}
We have the natural isomorphism
$${\vee}^r\W\cong\mathop{\oplus}\limits_{\sst h=0}^{\sst r}\!\U^{(r-h,h)}\equiv
\U^{(r,0)}\oplus\U^{(r-1,1)}
\oplus~\mdots~\oplus\U^{(1,r-1)}\oplus\U^{(0,r)}~,$$
characterized by the injections
\begin{multline*}
(u_1\ve\dots\ve u_h)\tn(\bl_1\ve\dots\ve\bl_k)\mapsto
(u_1,0)\ve\dots\ve(u_h,0)\ve(0,\bl_1)\ve\dots\ve(0,\bl_k)~,\\
u_1,\dots, u_h\in\U,~\bl_1,\dots,\bl_k\in\Ua\,,~0\leq h,k\leq r~.
\end{multline*}
\end{proposition}{\sc proof:~}
The above introduced maps indeed turn out to be injections
\hbox{$\smash{\mathop{\oplus}\limits_{\sst h=0}^{\sst r}\!
\U^{(r-h,h)}}\into{\vee}^r\W$}.
The stated isomorphism then follows from dimensional considerations.\qed

A certain extension of ${\vee}^r\W$ will turn out to yield
the most convenient setting for our formulation.
We introduce the further notation
$$\tilde\U{}^{(h,k)}\equiv{\vee}^{h}\Ua{\otimes}{\vee}^{k}\U~,$$
that is essentially a transposed bundle of $\U{}^{(k,h)}$,
and set
\begin{align*}
\W^{\{r\}}~&\equiv~{\vee}^r\W~\oplus~
\mathop{\oplus}\limits_{\sst h=1}^{\sst r-1}\!\td\U{}^{(r-h,h)}\equiv
\\[6pt]
&~\equiv {\vee}^r\W~\oplus~\Ua{\otimes}{\vee}^{r-1}\U~\oplus~
{\cdot}{\cdot}{\cdot}~\oplus~{\vee}^{r-1}\Ua{\otimes}\,\U~.
\end{align*}
Thus we may regard ${\vee}^r\W$ as the
sub-bundle of \hbox{$\W^{\{r\}}\subset{\otimes}^r\W$}
which is invariant under all transpositions
$$\td\U{}^{(h,r-h)}\longleftrightarrow\U^{(r-h,h)}~,\qquad 1\leq h\leq r-1~.$$

The standard Dirac map can be naturally extended
to be valued into endomorphisms of $\W^{\{r\}}$,
though the extension is no longer a Clifford map in general.
In fact $\g[\sY]$\,, for any \hbox{$\sY\in\H$},
yields maps
\begin{align*}
&{\vee}^{h}\U\tn{\vee}^{k}\Ua~\longrightarrow~
{\vee}^{h-1}\U\tn\Ua\tn{\vee}^{k}\Ua~,
\\[6pt]
&{\vee}^{h}\Ua\tn{\vee}^{k}\U~\longrightarrow~
{\vee}^{h-1}\Ua\tn\U\tn{\vee}^{k}\U~.
\end{align*}
An appropriate symmetrization then yields maps
$$\chg[\sY]:\U^{(h,k)}~\longrightarrow~\U^{(h-1,k+1)}~,\qquad
\chg[\sY]:\td\U{}^{(h,k)}~\longrightarrow~\td\U{}^{(h-1,k+1)}~,$$
with the coordinate expressions\footnote{
We use the convention that braces denoting index symmetrization
imply normalizing factorials, thus e.g.\ 
$\phi_{\{\!\sA\sB\}}\equiv\oh\,(\phi_{\sA\sB}+\phi_{\sB\sA})$\,.}
\begin{align*}
&(\chg[\sY]\Psi)\Ii{\sA_1\dots\sA_{h-1}}{\cB_1\dots\cB_{k+1}}=
\sqrt2\,\Psi\Ii{\sA_1\dots\sA_h}{\{\!\cB_1\dots\cB_k}\,
\sY^\IP_{\cB_{k+1}\}\sA_h}~,
\qquad \Psi\in\U^{(h,k)}~,
\\[6pt]
&(\chg[\sY]\Psi)\iI{\cA_1\dots\cA_{h-1}}{\sB_1\dots\sB_{k+1}}=
\sqrt2\,\Psi\iI{\cA_1\dots\cA_h}{\{\!\sB_1\dots\sB_k}\,
\sY^{\sB_{k+1}\}\cA_h}_\IP~,
\qquad \Psi\in\td\U{}^{(h,k)}~.
\end{align*}

Hence we get the sequence
\begin{multline*}
\U^{(r,0)}\stackrel{\chg[\ssY]}{\longrightarrow}\U^{(r-1,1)}~\mdots~
\stackrel{\chg[\ssY]}{\longrightarrow}\U^{(1,r-1)}
\stackrel{\chg[\ssY]}{\longrightarrow}\U^{(0,r)}\equiv\tilde\U^{(r,0)}
\stackrel{\chg[\ssY]}{\longrightarrow}
\\[6pt]
\stackrel{\chg[\ssY]}{\longrightarrow}\tilde\U^{(r-1,1)}~\mdots~
\stackrel{\chg[\ssY]}{\longrightarrow}\tilde\U^{(1,r-1)}
\stackrel{\chg[\ssY]}{\longrightarrow}\tilde\U^{(0,r)}\equiv\U^{(r,0)}~.
\end{multline*}

We now observe that for any \hbox{$\sY\in\H$} we have
$$\sY^\AAd\,\sY\!\!_{\sB\cA}=\oh\,g(\sY,\sY)\,\d^\sA_\sB~,\qquad
\sY^\AAd\,\sY\!\!_{\sA\cB}=\oh\,g(\sY,\sY)\,\d^\cA_\cB~,\qquad
\sY^\AAd\,\sY\!\!_\AAd=g(\sY,\sY)~,$$
and that \hbox{$\g[\sY]:\U\to\Ua$} and \hbox{$\g[\sY]:\Ua\to\U$}
are isomorphisms whenever $\sY$ is non-isotropic
(see remark in~\sref{ss:Two-spinors and Dirac spinors}).

\begin{proposition}
Let \hbox{$\sY\in\H$} be non-isotropic (\hbox{$g(\sY,\sY)\neq0$}).
Then the compositions
\begin{align*}
&(\chg[\sY])^r\equiv
\underbrace{\chg[\sY]\comp\mdots\comp\chg[\sY]}_{r~\text{factors}}~:~
\U^{(r,0)}\longrightarrow\U^{(0,r)}\equiv\tilde\U^{(r,0)}~,
\\[6pt]
&(\chg[\sY])^r\equiv
\underbrace{\chg[\sY]\comp\mdots\comp\chg[\sY]}_{r~\text{factors}}~:~
\tilde\U^{(r,0)}\longrightarrow\tilde\U^{(0,r)}\equiv\U^{(r,0)}~,
\end{align*}
are isomorphisms.
\end{proposition}{\sc proof:~}
Let \hbox{$\Psi\in\U^{(r,0)}$}.
Then by straightforward computations one finds
$$\bigl((\chg[\sY])^r(\Psi)\bigr)_{\cA_1\dots\cA_r}=
2^j\,\Psi^{\sA_1\dots\sA_r}\,\sY\!\!_{\sA_1\cA_1}\dots \sY\!\!_{\sA_r\cA_r}~,$$
and the like,
namely $(\chg[\sY])^r$ can be essentially viewed as the operation
of lowering all the indices through $\sY$ seen as a Hermitian metric.\qed

\begin{corollary}\label{corollary:seqmono}
Each non-isotropic \hbox{$\sY\in\H$} determines a monomorphism
$$\U^{(r,0)}\equiv{\vee}^r\U\into\W^{\{r\}}$$
by generating a sequence of isomorphic images of $\U^{(r,0)}$.
\end{corollary}

\begin{remark}~If \hbox{$g(\sY,\sY)=1$}\,, \hbox{$\Psi\in\U^{(r,0)}$}, then
$$(\chg[\sY])^r\comp(\chg[\sY])^r(\Psi)=\Psi~.$$
\end{remark}

\subsection{Generalised algebraic Dirac equation}
\label{ss:Generalised algebraic Dirac equation}

If $\th$ is a soldering form then we also have the transpose morphism
\hbox{$\th^*:\LL^{-1}\tn\H^*\to\TS\M$}.
Hence an element \hbox{$\sP\in\LL^{-1}\tn\H^*$}
such that \hbox{$g^\#(\sP,\sP)=m^2\in\LL^{-2}$}
can be regarded as a momentum of a particle of mass $m$\,.

In the symmetric higher-spin context
presented in~\sref{ss:Symmetric spinors}
we consider the obvious extension of the standard `algebraic Dirac equation'
\hbox{$\g[\sP]\psi=m\psi$} of standard electrodynamics, namely\footnote{
$\g[\sP]$ is a shorthand for $\g[g^\#(\sP)]$\,.}
$$\chg[\sP]\Psi=m\Psi~,\qquad \Psi\in\W^{\{2j\}}~,\quad2j\in\NN~.$$
We call this the \emph{generalised algebraic Dirac equation}.

We write
$$\Psi=\sum_{h=0}^{2j}\Psi^{(2j-h,h)}+\sum_{h=1}^{2j}\Psi'{}^{(2j-h,h)}~,
\qquad \Psi^{(2j-h,h)}\in\U^{(2j-h,h)}~,~~
\Psi'{}^{(2j-h,h)}\in\tilde\U^{(2j-h,h)}~.$$
Then the generalised algebraic Dirac equation reads
$$\begin{cases}
\tfrac1m\,\chg[\sP]\Psi^{(2j-h,h)}=\Psi^{(2j-h-1,h+1)}
\\[6pt]
\tfrac1m\,\chg[\sP]\Psi'{}^{(2j-h,h)}=\Psi'{}^{(2j-h-1,h+1)}
\end{cases}\quad 0\leq h\leq 2j-1~.$$

Recalling corollary~\ref{corollary:seqmono} we then see that
its solutions are characterized by
$$\Psi^{(2j,0)}\in\U^{(2j,0)}\equiv{\vee}^{2j}\U~,$$
which generates the values in the other sectors by repeated application
of the operator $\tfrac1m\chg[\sP]$\,.
Such solutions also fulfill
$$\begin{cases}
m^{-2j}\,(\chg[\sP])^{2j}\Psi^{(2j-h,h)}=\tilde\Psi'{}^{(2j-h,h)}
\\[6pt]
m^{-2j}\,(\chg[\sP])^{2j}\tilde\Psi^{(2j-h,h)}=\tilde\Psi'{}^{(2j-h,h)}
\end{cases}\quad 0\leq h\leq 2j~,$$
where a tilde denotes tensor product transposition
\hbox{$\U^{(2j-h,h)}
\stackrel{{\sim}}{\displaystyle\leftrightarrow}\tilde\U^{(h,2j-h)}$}.

The above condition is essentially the algebraic (momentum space)
version of the Joos-Weinberg equation for $\Psi$.
We note that this equation does not need the full extended space $\W^{\{2j\}}$,
but can be formulated, in a restricted setting, for
\hbox{$\Psi\in{\vee}^{2j}\U\oplus{\vee}^{2j}\Ua$}.
The latter space carries a \hbox{$2(2j\,{+}\,1)$}-dimensional representation
of the angular-momentum algebra,
which indeed corresponds to formulations
found in the literature~\cite{Weinberg1964i,DelgadoAcosta_etal2015}.

\subsection{Generalised Dirac equation and plane waves}
\label{ss:Generalised Dirac equation and plane waves}

In the context introduced
in~\sref{ss:Two-spinor soldering form and field theory},
the algebraic constructions
of~\sref{ss:Higher-spin extensions of the Dirac map}
and~\sref{ss:Generalised algebraic Dirac equation}
can be performed fiberwise.
We introduce a generalised Dirac operator,
acting on sections \hbox{$\Psi:\M\to\W^{\{2j\}}$}, as
$$\check\nasl\Psi\equiv\chg{}^a\,\na_a\Psi~,\qquad
\chg{}^a\equiv g^{ab}\,\th_b^\la\,\chg[\tau_\la]~.$$
We obtain the coordinate expressions
\begin{align*}
&(\check\nasl\Psi)\Ii{\sA_1\dots\sA_{h-1}}{\cB_1\dots\cB_{k+1}}=
\sqrt2\,\na_a\Psi\Ii{\sA_1\dots\sA_h}{\{\!\cB_1\dots\cB_k}\,
\th^{a}_{\cB_{k+1}\}\sA_h}~,
\\[6pt]
&(\check\nasl\Psi)\iI{\cA_1\dots\cA_{h-1}}{\sB_1\dots\sB_{k+1}}=
\sqrt2\,\na_a\Psi\iI{\cA_1\dots\cA_h}{\{\!\sB_1\dots\sB_k}\,
\th^{a\,\sB_{k+1}\}\cA_h}_\IP~.
\end{align*}

We now consider the special case of flat spacetime
with vanishing gauge field.\footnote{
More precisely we assume that the gauge field (a connection)
has vanishing curvature tensor, so that its components vanish in suitable frames.}
Thus \hbox{$\U\onto\M$} is now a trivial bundle as well as \hbox{$\TO\M\onto\M$},
and we replace $\na_a\Psi$ by $\de_a\Psi$\,.

Let \hbox{$\sP:\M\to\TS\M$} be constant, with \hbox{$g^\#(\sP,\sP)=m^2$}.
A section \hbox{$\Psi:\M\to\W^{\{2j\}}$} can be regarded as a `plane wave'
of positive energy if $\sP$ is future-pointing and there exists a fixed element
\hbox{$\underline\Psi\in\W^{\{2j\}}$} such that
$$\Psi(\sX)=\eO^{-\iO\,\bang{\sP\sX}}\,\underline\Psi~,$$
where $\sX$ is the `position-vector' in spacetime
with respect to any chosen `origin'.
Accordingly we get
$$\check\nasl\Psi=-\iO\,\sP\!_a\chg{}^a\Psi=-\iO\,\chg[\sP]\Psi~.$$
Thus $\Psi$ is a solution of the generalised Dirac equation
$$\iO\,\check\nasl\Psi=m\Psi~,$$
if and only if $\underline\Psi$ is a solution
of the generalised algebraic Dirac equation
$$\chg[\sP]\underline\Psi=m\underline\Psi~.$$
Namely, $\underline\Psi$ is generated from
$$\underline\Psi^{(2j,0)}\in\U^{(2j,0)}\equiv{\vee}^{2j}\U$$
by the repeated action of $\frac1m\chg[\sP]$\,.

Moreover such solutions also fulfill the Joos-Weinberg equation in the form
$$(-\iO)^{2j}\,\check\nasl{}^{2j}\Psi=m^{2j}\,\Psi~,$$
which can be also formulated in a restricted setting for a  field
\hbox{$\M\to{\vee}^{2j}\U\oplus{\vee}^{2j}\Ua$}.

\subsection{Lagrangian}
\label{ss:Lagrangian}

The Dirac map acts on \hbox{$\Wl\cong\Ul\oplus\Uc$} by standard
linear map transposition.
Dual constructions are then straightforward.
In particular for all \hbox{$\sY:\M\to\H$} we get a morphism
\hbox{$\chg[\sY]:\Wl{}^{\!\{2j\}}\to\Wl{}^{\!\{2j\}}$},
with the coordinate expressions
\begin{align*}
&(\chg[\sY]\bar\Psi)\iI{\sA_1\dots\sA_{h-1}}{\cB_1\dots\cB_{k+1}}=
\sqrt2\,\bar\Psi\!\iI{\sA_1\dots\sA_h}{\{\!\cB_1\dots\cB_k}\,
\sY_{\phantom{A_h}}^{\cB_{k+1}\}\sA_h}~,
\\[6pt]
&(\chg[\sY]\bar\Psi)\Ii{\cA_1\dots\cA_{h-1}}{\sB_1\dots\sB_{k+1}}=
\sqrt2\,\bar\Psi\Ii{\cA_1\dots\cA_h}{\{\!\sB_1\dots\sB_k}\,
\sY\!\!_{\sB_{k+1}\}\cA_h}^\IP~.
\end{align*}

Besides \hbox{$\Psi:\M\to\W^{\{2j\}}$} we consider
an independent field \hbox{$\bar\Psi:\M\to\Wl{}^{\!\{2j\}}$}.
We then write down the Lagrangian density
\hbox{$\Lcal_\Psi=\ell\,\dO^4\xx$}\,, where
$$\tfrac1{|\th|}\,\ell\equiv
\ih\,\bigl(\bang{\bar\Psi,\check\nasl\Psi}-\bang{\check\nasl\bar\Psi,\Psi}\bigr)
-m\,\bang{\bar\Psi,\Psi}~.$$

The field equations for the couple $(\Psi,\bar\Psi)$
can be expressed in the form
$$\begin{cases}
-\iO\,\check\nasl\bar\Psi-m\bar\Psi-\ih\bar\Psi\,\chg[\breve T]=0~,
\\[6pt] \phantom{-}
\iO\,\check\nasl\Psi-m\Psi-\ih\chg[\breve T]\Psi=0~,
\end{cases}$$
where \hbox{$\breve T=\breve T_a\,\dx^a\equiv T\Ii{b}{ab}\,\dx^a$}
is the 1-form associated with the torsion
(\sref{ss:Two-spinor soldering form and field theory}).
Formally these look exactly as the standard Dirac equations
in tetrad-affine gravity,
apart for the apperance of the generalised Dirac map and Dirac operator.

\smallbreak\noindent{\sc computation.}
We can derive the field equations by means
of the `covariant-differential' approach,
in which the fields' components and their covariant derivatives
are to be regarded as \emph{independent variables}~\cite{C16e}.
We introduce $\W^{\{2j\}}$-valued
and $\Wl{}^{\!\{2j\}}$-valued exterior $r$-forms $\Pi^{\sst(r)}$ on $\M$,
\hbox{$r=0,1$},
characterised (by some abuse of language) as
$$\Pi^{\sst(0)}\equiv\dde{\Lcal_\Psi}\Psi~,\qquad
\bar\Pi^{\sst(0)}\equiv\dde{\Lcal_\Psi}{\bar\Psi}~,\qquad
\Pi^{\sst(1)}\equiv\dde{\Lcal_\Psi}{(\nabla\Psi)}~,\qquad
\bar\Pi^{\sst(1)}\equiv\dde{\Lcal_\Psi}{(\nabla\bar\Psi)}~.$$
We obtain
\begin{align*}
& \Pi^{\sst(0)}=\bigl(-\ih\,\check\nasl\bar\Psi-m\bar\Psi\bigr)\,|\th|\,\dO^4\xx~,
&& \Pi^{\sst(1)}=\ih\,|\th|\,\bar\Psi\,\chg{}^a\tn\dx_a~,
\\[6pt]
& \bar\Pi^{\sst(0)}=\bigl(\ih\,\check\nasl\Psi-m\Psi\bigr)\,|\th|\,\dO^4\xx~,
&& \bar\Pi^{\sst(1)}=-\ih\,|\th|\,\chg{}^a\Psi\,\tn\dx_a~,
\end{align*}
where \hbox{$\dx_a\equiv i(\de\xx_a)\dO^4\xx$}\,.
Then the field equations can be written as
$$\begin{cases}
\Pi^{\sst(0)}-\dO_{\Cs}\Pi^{\sst(1)}=0~,
\\[6pt]
\bar\Pi^{\sst(0)}-\dO_{\Cs}\bar\Pi^{\sst(1)}=0~.
\end{cases}$$
Here $\dO_{\Cs}$ denotes the \emph{covariant differential}
of vector-valued forms.
For \hbox{$r=1$} this can be also expressed as
$$\dO_{\Cs}\Pi^{\sst(1)}=\codiv\Pi{}^{\sst(1)}+\tau\we\Pi^{\sst(1)}~,\qquad
\dO_{\Cs}\bar\Pi^{\sst(1)}=\codiv\bar\Pi{}^{\sst(1)}+\tau\we\bar\Pi^{\sst(1)}~,$$
where $\codiv{}$ denotes the covariant divergence operator.
Since the covariant derivatives of $\th$ and $\chg$ vanish
we also get
$$\codiv\Pi^{\sst(1)}=\ih\,\check\nasl\bar\Psi~,\qquad
\codiv\bar\Pi{}^{\sst(1)}=-\ih\,\check\nasl\Psi~,$$
whence the stated result follows.\qed

\bigbreak
In Lagrangian field theories one can consider
the notion of \emph{canonical energy-tensor}
associated with a certain field.
Evaluated through the field, the canonical energy-tensor is a section
\hbox{$\Ucal:\M\to\TO\M\tn\weu3\TS\M$}.
In non-trivial bundles $\Ucal$ can be introduced as a geometrically well-defined
tensor field with the intervention of a suitable connection;
possibly, the latter can be the gauge field itself~\cite{C16c}.
In the case of the theory under consideration we find
\begin{align*}
\Ucal\Ii ab&=\ell\,\d\Ii ab
-\Bang{\dde{\Lcal_\Psi}{(\na_a\Psi)}\,,\,\na_b\Psi}
-\Bang{\dde{\Lcal_\Psi}{(\na_a\bar\Psi)}\,,\,\na_b\bar\Psi}=
\\[8pt]
&=\ell\,\d\Ii ab
-\ih\,|\th|\,\Bang{\bar\Psi\,\chg{}^a\,,\,\na_b\Psi}
+\ih\,|\th|\,\Bang{\bar\Psi\,,\,\chg{}^a\na_b\Psi}~.
\end{align*}
This expression is a generalization
of the canonical energy-tensor for the Dirac field:
the essential difference consists in the fact that
one has now a sum over fiber contractions
in all sectors of $\W^{\{2j\}}$ and $\Wl{}^{\!\{2j\}}$.

\subsection{Gauge field interaction}
\label{ss:Gauge field interaction}

In~\sref{ss:Lagrangian} the connection yielding
the covariant derivatives $\nabla\Psi$ and $\nabla\bar\Psi$
is assumed to contain the gauge field
as well as a purely gravitational contribution,
according to the setting presented
in~\sref{ss:Two-spinor soldering form and field theory}.
At that level, no essential formal change is needed in order to include
non-Abelian gauge fields.
Entering further detail about gauge field interaction,
the most obvious procedure consists in
replacing the two-spinor bundle $\U$ with \hbox{$\U'\cong\U\tn\F$},
where \hbox{$\F\onto\M$} is a complex vector bundle endowed with
a Hermitian structure.
With no loss of generality, the gauge part
of the considered connection $\Cs'$ of $\U'$
can be completely attributed to a linear Hermitian connection $\k$ of $\F$,
leaving the gravitational contribution associated with $\U$ only.
Accordingly we write the components of $\Cs'$ as
$$\Cs'\!\iIi a{\sA i}{\sB j}\equiv
\Cs\!\iIi a\sA\sB\,\d\Ii ij+\d\Ii\sA\sB\,\k\iIi aij~,$$
where \hbox{$\Cs\!\iIi a\sA\sA=0$}
namely $\Cs$ is a `purely gravitational' connection of $\U$.

When we deal with arbitrary spin fields, we have further choices
as to how these are to interact with the gauge fields.
Let us consider two possibilities.

\smallbreak\noindent
{\bf a)}~We may construct higher-spin bundles $\W'{}^{\{2j\}}$
by using $\U'$ instead of $\U$.
Taking into account the isomorphism \hbox{$\F\cong\Fa$}
determined by the assumed Hermitian structure, we get
$$\U'{}^{(h,k)}\cong
{\vee}^{h}\U'\tn{\vee}^{k}\Uc{}'{}^\lin\tn{\vee}^{h}\F\tn{\vee}^{k}\F$$
and the like.
The extension $\chg$ of the Dirac map naturally acts on these bundles,
so that the arguments
of~\sref{ss:Generalised algebraic Dirac equation}|\ref{ss:Lagrangian}
still apply, essentially unchanged.
The gauge field interaction is then determined
by the appropriate tensor power of $\Cs'$,
hence the interaction with the sector $\U'{}^{(h,2j-h)}$
may depend on $h$\,.
In the Abelian case, however, we get the same gauge field interaction
for all sectors of spin $j$\,,
with the interaction charge turning out to be the $2j$-th power of the
charge for spin one-half.

\smallbreak\noindent
{\bf b)}~A somewhat simpler theory can be considered by setting
$$\W'{}^{\{2j\}}\equiv\W^{\{2j\}}\tn\F~,$$
thus allowing the gauge field interaction to be the same
in all spin sectors.
\bigbreak

Whatever scheme we choose for describing the interaction between
the gauge field and arbitrary-spin fields,
the gauge Lagrangian \hbox{$\Lcal\spec{gauge}=\ell\spec{gauge}\,\dO^4\xx$}
is assumed to be of the usual type
$$\ell\spec{gauge}=\oq\,g^{ac}\,g^{bd}\,\rho\iIi{ab}ij\,\rho\iIi{cd}ji\,|\th|~,$$
where
$$\rho\equiv-\dO_\k\k:\M\to\weu2\TS\M\tn\End\F$$
is the curvature tensor of the connection $\k$\,.
Using again the aforementioned covariant-differential approach~\cite{C16e},
the `second Maxwell equation' is readily seen to be
$$\oh\,\dO_\k({*}\dO_\k\k)=\Jcal~,$$
where the `current' in the right-hand side can be expressed,
in a loose notation, as
$$\Jcal=-\dde{\ell_\Psi}{(\nabla\Psi)}\,\dde{(\nabla\Psi)}{\k}
+\dde{\ell_\Psi}{(\nabla\bar\Psi)}\,\dde{(\nabla\bar\Psi)}{\k}~.$$
The explicit form of $\Jcal$ depends on the chosen scheme.
In case {\bf a)} one gets a somewhat intricate expression,
though straightforwardly computable if needed.
In case {\bf b)}, on the other hand, one gets the much simpler expression
$$\Jcal=\Pi^{\sst(1)}\tn\Psi-\bar\Psi\tn\bar\Pi^{\sst(1)}~.$$

Finally we note that the canonical energy-tensor for the gauge field
has the expression
$$(\Ucal\spec{gauge})\Ii ab=
\ell\spec{gauge}\,\d\Ii ab
+2\,\dde{\ell\spec{gauge}}{\rho\iIi{ac}ij}\,\rho\iIi{bc}ij=
\bigl(\oq\,\rho\Ii{cd\,i}j\,\rho\iIi{cd}ji\,\d\Ii ab
-\rho\Ii{ac\,i}j\,\rho\iIi{bc}ji\bigr)\,|\th|~,$$
that is the same as in a generic gauge field theory.

\subsection{Further spinor field types}
\label{ss:Further spinor field types}

The general procedure for obtaining a first-order theory,
sketched in~\sref{ss:Higher-spin extensions of the Dirac map},
can be adapted to other field types besides symmetric spinors.

A section \hbox{$\sV:\M\to\U{\otimes}\,\Uc$} can be seen as a `complexified'
vector field, of spin \hbox{$j=1$}\,.
We may write its coordinate expression as \hbox{$\sV^\la\tau_\la$}\,,
namely its components have a spacetime index.
In covariant form it may be used to represent a deformation of a gauge field.
Extensions \hbox{$\sV':\M\to(\U\oplus\Ua)\tn\Uc$}
and \hbox{$\sV'':\M\to\U\tn(\Uc\oplus\Ul)$}
are acted upon by operators \hbox{$\nasl{}'\equiv\nasl\!_{\sst(1)}$}
and \hbox{$\nasl{}''\equiv\nasl\!_{\sst(2)}$}
(see~\sref{ss:Higher-spin extensions of the Dirac map}).
Then we obtain the first-order field equations
$$\iO\,\nasl{}'\sV'=m\,\sV'~,\qquad -\iO\,\nasl{}''\sV''=m\,\sV''~.$$

We note that $\sV'$ and $\sV''$ are valued into bundles
that are mutually conjugate up to tensor transposition.
If $\sV'$ and $\sV''$ are mutually conjugate-transposed,
then the two above equations turn out to be actually equivalent.
In flat spacetime we get plane wave solutions that are determined
by their values in the `main sector' \hbox{$\H\equiv\HO(\U{\otimes}\,\Uc)$}\,.

Anti-symmetric spinors can be treated similarly to symmetric spinors.
In particular, since we have the natural isomorphism
$$\weu2\W\cong\weu2\U~\oplus~\U\!\tn\Ua~\oplus~\weu2\Ua~,$$
we also consider the extension
$$\W^{[2]}\equiv\weu2\U~\oplus~\U\!\tn\Ua~\oplus~\weu2\Ua
~\oplus~\Ua\!\tn\U~.$$
Then, by using a suitable extension of the Dirac map,
for all \hbox{$\sY:\M\to\H$} we obtain the sequence
$$\weu2\U\stackrel{\hag[\ssY]}{\longrightarrow}\U\tn\Ua
\stackrel{\hag[\ssY]}{\longrightarrow}\weu2\Ua
\stackrel{\hag[\ssY]}{\longrightarrow}\Ua\tn\U
\stackrel{\hag[\ssY]}{\longrightarrow}\weu2\U~.$$
A first-order field equation of Dirac type can then be introduced;
in the flat spacetime case this admits plane wave solutions that are
characterized by their value in the main sector $\weu2\U$.

Next we consider a spin-2 field,
thought of representing deformations of the metric.
In contravariant form this can be described as a section
\hbox{$\M\to{\vee}^2\H$}.
By standard spinor algebra methods it is not difficult to show
that we have the natural decomposition
$${\vee}^2\!\H\cong
\HO({\vee}^2\U\tn{\vee}^2\Uc)~\oplus~\HO(\weu2\U\tn\weu2\Uc)~.$$
For various reasons it will be convenient to consider a `complexified'
version, namely a field
$$G\equiv\check G+\hat G
:\M\to({\vee}^2\U\tn{\vee}^2\Uc)~\oplus~(\weu2\U\tn\weu2\Uc)\cong
\CC\tn{\vee}^2\!\H~,$$
with components
\hbox{$G^{\sA\sB\cA\cB}\equiv G^{\sA\cA\sB\cB}\equiv
\check G^{\sA\sB\cA\cB}+\hat G^{\sA\sB\cA\cB}$}
where
$$\check G^{\sA\sB\cA\cB}=G^{\{\!\sA\sB\}\{\!\cA\cB\}}~,\qquad
\hat G^{\sA\sB\cA\cB}=
\oq\,(\e_{\sC\sD}\,\be_{\cC\cD}\,G^{\sC\cC\sD\cD}\,)\,
\e^{\sA\sB}\,\be^{\cA\cB}~.$$
Note that $\hat G$, being proportional to the natural contravariant metric
$\e^\#\tn\be^\#$ of $\H^*$,
can be regarded as a `dilatonic' field.

The terms $\check G$ and $\hat G$ can be treated as independent fields,
and the first-order theory introduced in previous sections
can be adapted to both cases.
We give a succinct description, starting from $\check G$.
Consider the extensions
\begin{align*}
&\check G':\M\to\W^{\{2\}}\tn{\vee}^2\Uc~,
&&\chg{}'\equiv\chg\tn\id:\H\to\End\bigl(\W^{\{2\}}\tn{\vee}^2\Uc\bigr)~,
\\[6pt]
&\check G'':\M\to{\vee}^2\U\tn\Wc^{\{2\}}~,
&&\chg{}''\equiv\id\tn\chg:\H\to\End\bigl({\vee}^2\U\tn\Wc^{\{2\}}\bigr)~.
\end{align*}

Then we obtain `Dirac' operators \hbox{$\check\nasl{}'\equiv\chg'{}^a\na_a$}
and \hbox{$\check\nasl{}''\equiv\chg''{}^a\na_a$}\,,
yielding the first-order field equations
$$\iO\,\check\nasl{}'\check G'=m\,\check G'~,\qquad
-\iO\,\check\nasl{}''\check G''=m\,\check G''~.$$
As in the case of a spin-$1$ field,
if $\check G'$ and $\check G''$ are mutually conjugate-transposed
then the two introduced equations turn out to be actually equivalent.
In flat spacetime we get plane wave solutions that are determined
by their values in the `main sector'
\hbox{$\HO({\vee}^2\U\tn{\vee}^2\Uc)$}.

The field \hbox{$\hat G:\M\to\weu2\U\tn\weu2\Uc$}
can be treated in a similar way,
starting from the previously sketched anti-symmetric spinor version.



\begin{thebibliography}{40}
%
\bibitem{BargmannWigner48}
V.\ Bargmann and E.P.\ Wigner:
`Group theoretical discussion of relativistic wave equations',
Proceedings of the National Academy of Sciences of the United States of America
{\bf 34} N.5 (1948), 211--23.
%
\bibitem{BarthChristensen83}
N.H.\ Barth and S.M.\ Christensen:
`Arbitrary spin field equations on curved manifolds with torsion',
J.\ Phys.\ A {\bf 16} (1983), 543--563.
%
\bibitem{BlaineLawsonMichelsohn89}
H.\ Blaine Lawson Jr. and M.-L.\ Michelsohn:
\emph{Spin Geometry}, Princeton University Press, Princeton (1989).
%
\bibitem{C98}
D.\ Canarutto:
`Possibly degenerate tetrad gravity and Maxwell-Dirac fields',
J.\ Math.\ Phys.\ {\bf 39}, N.9 (1998), 4814--4823.
%
\bibitem{C00b}
D.\ Canarutto:
`Two-spinors, field theories and geometric optics in curved spacetime',
Acta Appl.\ Math.\ {\bf 62} N.2 (2000), 187--224.
%
\bibitem{C07}
D.\ Canarutto:
`{}``Minimal geometric data'' approach to
Dirac algebra, spinor groups and field theories',
Int.\ J.\ Geom.\ Met.\ Mod.\ Phys., {\bf 4} N.6, (2007), 1005--1040.\\
arXiv:math-ph/0703003.
%
\bibitem{C10a}
D.\ Canarutto:
`Tetrad gravity, electroweak geometry and conformal symmetry',
Int.\ J.\ Geom.\ Met.\ Mod.\ Phys., {\bf 8} N.4 (2011), 797--819;
arXiv:1009.2255v1 [math-ph].
%
\bibitem{C12a}
D.\ Canarutto: 
Positive spaces, generalized semi-densities and quantum interactions. 
J.\ Math.\ Phys. {\bf53} (3), 032302 (2012).
%
\bibitem{C14}
D.\ Canarutto:
`Two-spinor geometry and gauge freedom.',
Int.\ J.\ Geom.\ Met.\ Mod.\ Phys., {\bf 11} (2014),
DOI: http://dx.doi.org/10.1142/S0219887814600160;\\
arXiv:1404.5054 [math-ph].
%
\bibitem{C15b}
D.\ Canarutto: 
`Natural extensions of electroweak geometry and Higgs interactions',
Ann.\ H.\ Poincar\'e {\bf16} N.11 (2015), 2695--2711;
arXiv:1407.4312 [math-ph]
%
\bibitem{C16c}
D.\ Canarutto: 
`Overconnections and the energy-tensors of gauge and gravitational fields',
J.\ Geom.\ Phys.\ {\bf106} (2016), 192--204.
%
\bibitem{C16e}
D.\ Canarutto: 
`Covariant-differential formulation of Lagrangian field theory',\\
arXiv:1607.03864 [math-ph]
%
\bibitem{Canarutto2018a}
D.\ Canarutto: 
`On the notions of energy tensors in tetrad-affine gravity',\\
Gravitation~\&~Cosmology {\bf24} N.\,2 (2018), \emph{to appear}\\
arXiv:1708.08109 [math-ph]
%
\bibitem{DabrowskiPercacci86}
L.\ Dabrowski and R.\ Percacci:
`Spinors and diffeomorphisms', 
Comm.\ Math.\ Phys.\  {\bf106} (1986), 691--704.
%
\bibitem{DelgadoAcosta_etal2015}
E.G.\ Delgado Acosta, V.M.\ Banda Guzm\'{a}n and M.\ Kirchbach:
`Bosonic and fermionic Weinberg-Joos \hbox{$(j,0)\oplus(0,j)$}
states of arbitrary spins as Lorentz tensors or tensor-spinors
and second-order theory',
Eur.\ Phys.\ J.\ A (2015) 51:~35\\
DOI 10.1140/epja/i2015-15035-x.
%
\bibitem{Dvoeglazov2003}
V.V.\ Dvoeglazov:
`Generalizations of the Dirac equation and the modified
Bargmann-Wigner formalism',
Hadronic J.\ {\bf26} (2003), 299--325.\\
arXiv:02081592v2 [hep-th].
%
\bibitem{FatibeneFrancaviglia98}
L.\ Fatibene and M.\ Francaviglia:
`Deformations of spin structures and gravity', in
\emph{Gauge theories of gravitation} (Jadwisin, 1997).
Acta Phys.\ Polon.\ B~{\bf29} No~4 (1998), 915--928.
%
\bibitem{HehlHeydeKerlickNester76}
F.W.\ Hehl, P.\ von der Heyde, G.\ D.\ Kerlick, and J.\ M.\ Nester:
`General relativity with spin and torsion: Foundations and prospects',
Reviews of Modern Physics {\bf48} n.3 (1976), 393-416.
%
\bibitem{Henneaux1978}
M.\ Henneaux:
`On geometrodynamics with tetrad fields',
Gen.\ Rel.\ Grav.\ {\bf9} (1978), 1031--1045.
%
\bibitem{HuangRuanWuZheng2002}
Huang Shi-Zhong, Ruan Tu-Nan, Wu Ning and Zheng Zhi-Peng:
`Wavefunctions for Particles with Arbitrary Spin',
Commun.\ Theor.\ Phys.\ (Beijing, China) {\bf 37} (2002), 63--74.
%
\bibitem{Hurley1971}
W.\ Hurley:
`Relativistic wave equations for particles with arbitrary spin',
Phys.\ Rev. D {\bf 4} N.12 (1971), 3605--3616.
%
\bibitem{Hurley1972}
W.\ Hurley:
`Consistent description of higher-spin fields',
Phys.\ Rev. Lett.\ {\bf 29} N.21 (1972), 1475--1477.
%
\bibitem{JMV10}
J.\ Jany{\v s}ka, M.\ Modugno and R.\ Vitolo:
`An algebraic approach to physical scales',
Acta Appl.\ Math.\ {\bf 110} N.3 (2010), 1249--1276;
arXiv:0710.1313v1.
%
\bibitem{Jeffery78}
E.A.\ Jeffery:
`Component Minimization of the Bargmann-Wigner Wavefunction',
Aust.\ J.\ Phys.\ {\bf 31} (1978), 137--49.
%
\bibitem{Joos1962}
H.\ Joos,
`Zur Darstellungtheorie der inhomogenen Lorentzgruppe
als Grundlage quantenmechaniscer Kinematik',
Fortschritte der Physik {\bf10} (1962), 65--146.
%
\bibitem{Kaparulin_etal2013}
D.S.\ Kaparulin, S.L.\ Lyakhovic and S.S.\ Sharapov:
`Lagrangian anchor for Bargmann-Wigner equations',
Geometric Methods in Physics. XXXI Workshop 2012.
Trends in Mathematics 2013, 119--126;
arXiv:1210.2134v2 [math-ph].
%
\bibitem{LorenteRodriguez1983}
M.\ Lorente and M.A.\ Rodriguez:
`A Lagrangian of BargmannÐWigner equations for massive particles of spin 2',
J. Math.\ Phys.\ {\bf 24} N.12 (1983), 2823--2827.
%
\bibitem{PR84}
R.\ Penrose and W.\ Rindler:
\emph{Spinors and space-time.
{I}: Two-spinor calculus and relativistic fields},
Cambridge University Press, Cambridge (1984).
%
\bibitem{PR88}
R.\ Penrose and W.\ Rindler:
\emph{Spinors and space-time.
{II}: Spinor and twistor methods in space-time geometry},
Cambridge University Press, Cambridge (1988).
%
\bibitem{Percacci86}
R.\ Percacci:
`Gauge group of gravity, spinors and anomalies',
Int.\ J.\ Theor.\ Phys.\ {\bf25}~No~5 (1986), 493--507.
%
\bibitem{Shay1968}
D.\ Shay:
`A Lagrangian formulation of the Joos-Weinberg wave equations
for spin-$s$ particles',
Il Nuovo Cimento Vol.~LVII\,A, N.2 (1968), 210--218.
%
\bibitem{Weinberg1964i}
S.\ Weinberg:
`Feynman rules for any spin',
Phys.\ Rev. {\bf 133} N.5B (1964), 1318--1332.
%
\bibitem{Weinberg1964ii}
S.\ Weinberg:
`Feynman rules for any spin. II. Massless particles',
Phys.\ Rev. {\bf 134} N.4B (1964), 882--895.
%
\bibitem{Weinberg1969}
S.\ Weinberg:
`Feynman rules for any spin. III',
Phys.\ Rev. {\bf 181} N.5 (1969), 1893--1899.
%
\bibitem{Trautman06}
A.\ Trautman: `Einstein-Cartan theory',
in \emph{Encyclopedia of Mathematical Physics}, Vol.\ 2, edited by
J.-P.\ Fran\c{c}oise, G.L.\ Naber and Tsou S.T.,
Elsevier, Oxford (2006), 189--195.
%
\end{thebibliography}
\end{document}